\documentstyle[preprint,aps]{revtex}
\tightenlines
\begin{document}
\draft

\title{Method of integral transforms for calculating few--body reactions}

\author{V.D. Efros$^{1)}$ \thanks{supported by 
INFN and the Russian 
Foundation for Basic Research (grant no. 97-02-17003).},
W. Leidemann$^{2)}$, and G. Orlandini$^{2)}$}
\address{
1)Russian Research Centre "Kurchatov Institute",
Kurchatov Square 1,\\ 123182 Moscow, Russia\\
2)Dipartimento di Fisica,
Universit\`a di Trento, I-38050 Povo (Trento), Italy \\and
Istituto Nazionale di Fisica Nucleare, Gruppo Collegato di Trento}
\date{}

\maketitle

\begin{abstract}
A non--conventional approach to calculating reactions in quantum
mechanics is presented. Reaction observables are
obtained with bound state calculation techniques. The accuracy of the
method  to calculate few--nucleon response functions is discussed.
\end{abstract}
\bigskip

\vfill\eject

\section{INTRODUCTION}

The purpose of the method discussed is to solve reaction problems 
with bound state calculation techniques. The
approach can be applied to both perturbation and strong interaction
induced inclusive and exclusive processes.
Below the case of inclusive perturbation induced reactions is
considered.

The general procedure was introduced in 1985 \cite{Ef85}. 
The originally proposed Stieltjes transform led to stability 
problems \cite{ELO93}. In Ref. \cite{ELO94} it was found that a 
Lorentzian kernel allowed to cure the instabilities. Accurate
results on three- and four-nucleon response
functions were obtained with this kernel
\cite{Sara95,ELO97,ELO97b,ELO97c}.

In Sec. 2 the procedure is outlined. In Sec. 3 the stability issue
is explained and the inversion technique is described. 
In Sec. 4 the choice of the transform kernel is discussed. 
In Sec. 5 the accuracy of the method in the "exactly
solvable" $d(e,e')np$ problem is demonstrated, and its features
in the calculations of the three- and four-nucleon response 
functions are discussed.

\section{OUTLINE OF THE METHOD}

We need to obtain the response functions of the form
\begin{equation}
R(\epsilon)=\int df |\langle\Psi_f|\hat{O}|
\Psi_0\rangle|^2\delta(E_f-E_0-\epsilon)
\end{equation}
starting from a few--body Hamiltonian.
Here $\epsilon$ is the excitation energy, $\Psi_0$ is the ground state
wave function, $\Psi_f$ is a complete set of the final state wave
functions,
and $\hat{O}$ is a transition operator. In general the continuum wave
functions $\Psi_f$
include infinite numbers of various channels with three or more
fragments.
The number of these functions
is infinite as well. Therefore the calculation of $R$
via direct use of Eq. (1) would in general be impractical.
Complications may occur even at very low
energy due to the Coulomb interaction in channels with more than two
fragments.
The present method circumvents the above problems due to the fact that
the continuum wave functions $\Psi_f$ do not enter the calculation at
all.

The method proceeds in two steps. At the first step an integral
transform of $R$  with a smooth kernel $K$
\begin{equation}
\Phi(\sigma)=\int K(\sigma,\epsilon)R(\epsilon)d\epsilon,
\end{equation}
is calculated instead of $R$ itself.  Using sum--rule techniques one
can see that
\begin{equation}
\Phi(\sigma)=\langle\Psi_0|\hat{O}^{\dag} K(\sigma,\hat{H}-E_0)\hat{O}|
\Psi_0\rangle,
\end{equation}
$\hat{H}$ and $E_0$ being the Hamiltonian and the ground state energy,
respectively. Expression (3) is evaluated with {\em bound--state
techniques} as described below. As a second step Eq. (2)
is considered as an integral equation to invert and thus one gets $R(\epsilon)$
from $\Phi(\sigma)$.

\section{INVERSION OF THE TRANSFORM AND ISSUE OF STABILITY}

Let us suppose that the transform (3) is evaluated and consider the
subsequent inversion of Eq. (2). As it is known solutions to such type of 
equations are unstable with respect to high frequency oscillations. 
Indeed, let us add an oscillating
increment $\Delta R$ to the right--hand side of Eq. (2). The transform
will get the increment $\Delta \Phi$, and if the oscillations are rapid
enough $\Delta \Phi$ may be small even if the oscillation
amplitudes are not small. The quantity $\Delta \Phi$ might become less
than possible errors in the numerical calculation of the transform
and the corresponding rapidly oscillating
increment $\Delta R$ could not then be discriminated. 
It is known that a regularization procedure can cure the
instability. The mathematical justification of such procedures can be
found e.g. in Ref. \cite{TiA73}. The regularization 
suppresses rapid oscillations.
The higher  an accuracy in $\Phi$ is the higher might be the frequency level 
at which the suppression occurs and the finer are the details of a true 
response that can be reproduced with an approximate solution. 
A stability of the solution
with respect to changes in the regularization parameter would then
indicate that the high frequencies suppressed are insignificant for
reproducing a response.

The inversion procedure in the calculations
\cite{Ef85,ELO93,ELO94,Sara95,ELO97,ELO97b,ELO97c}
was the following.
The response was sought for in the form
\begin{equation}
R(\epsilon)=\sum_{n=1}^N c_n\chi_n(\epsilon,\alpha)\,,   
\label{eq:exp}
\end{equation}
where $\chi_n$ are known functions including nonlinear parameters
$\alpha$. 
If one substitutes this form into the right hand side of Eq. (2)
one has
\begin{equation}
\Phi(\sigma)=\sum_{n=1}^N c_n\tilde{\chi}_n(\sigma,\alpha)\,,
\end{equation}
where the $\tilde{\chi}_n(\sigma,\alpha)$ are the transformed 
basis functions. The parameters $c_n$ and $\alpha$ are obtained 
fitting the transform of Eq. (5) to the known results obtained 
by Eq. (3) at many $\sigma$ points. The number of functions $N$ 
plays the role of the regularization parameter and is chosen 
within the above mentioned stability region.

Features of a response known beforehand such as its low
energy behavior and possible information on narrow levels
could easily be incorporated into the set
$\{\chi_n\}$. This increases the accuracy of the inversion. The
longitudinal $(e,e')$ few--body responses behave in a rather peculiar
way at small $\epsilon$.  This is caused by the fact that lower
multipole contributions to the continuum part of the responses can have
maxima in the threshold region, while higher multipoles 
exhibit maxima only in the quasielastic peak region. 
We applied the method separately to
those lower multipole responses and to the sum of all others
retained \cite{ELO94,ELO97}. This substantially 
increased the accuracy of
the results at low energy.  The reason is that for simple functions
$\chi_n$ one can better describe the
behavior of those pieces of the response function with sufficiently low
$N$ values in Eq. (\ref{eq:exp}) than the total response function.

\section{THE CHOICE OF THE TRANSFORM KERNEL}

At a given accuracy in the transform $\Phi$ the resulting response
$R$ will be most accurate if the kernel is chosen as
"narrow" as possible. Indeed, only changes in $R$ occurring at the
intervals
$\Delta\epsilon$ smaller than the range of the kernel may be hard to
resolve
since they may tend to cancel
at integrating in Eq. (2).

The following choice proved to be efficient \cite{ELO94}.
Let the transform variable $\sigma$ be complex and consider the kernel
of the Lorentz form
\begin{equation}
K(\sigma=-\sigma_R+i\sigma_I,\epsilon)=
\frac{1}{(\sigma^*+\epsilon)(\sigma+\epsilon)}
=\frac{1}{[(\sigma_R-\epsilon)^2+\sigma_I^2]},     \label{eq:ker}
\end{equation}
$\sigma_I$ being "sufficiently small". According to Eq. (3) this leads
to
the transform
\begin{equation}
\Phi(\sigma)=\langle\Psi_0|\hat{O}^{\dag}\frac{1}{H-E_0+\sigma^*}
\frac{1}{H-E_0+\sigma}\hat{O}|\Psi_0\rangle. \label{eq:lor}
\end{equation}
Eq. (\ref{eq:lor}) can be written as
\begin{equation}
\Phi(\sigma)=\langle\tilde{\Psi}(\sigma)|\tilde{\Psi}(\sigma)\rangle\,,
\end{equation}
where $\tilde{\Psi}$ is the unique solution to the 
Schr\"odinger--like equation with a {\em source},
\begin{equation}
(H-E_0-\sigma_R+i\sigma_I)\tilde{\Psi}=Q,
\,\,\,\,\,\,\,\,\,\,\,\,\,\,\,\, Q=\hat{O}\Psi_0.
\label{eq:equ}
\end{equation}
Since $\Phi$ and thus
$\langle\tilde{\Psi}|\tilde{\Psi}\rangle$ does exists, $\tilde{\Psi}$
is localized! Hence bound--state type methods can be applied to solve
Eq. (\ref{eq:equ}).

\section{CALCULATIONS OF FEW--NUCLEON RESPONSES}

The accuracy of the method was tested with the two--body longitudinal
$d(e,e')$ response. To approach the situation with A=3
and 4, several percent errors were {\em artificially} introduced into
the transform, see \cite{ELO94}. Fig. 1 demonstrates that even in presence of
the errors the response obtained practically coincides with the exact
one calculated in the conventional way.

Longitudinal $(e,e')$ response functions \cite{Sara95,ELO97}
and total photoabsorption
cross sections \cite{ELO97b,ELO97c}
were studied for three- and four-nucleon cases.
In \cite{Sara95} Reid and Bonn realistic NN forces were employed. Eq.
(\ref{eq:equ}) was cast into the form of inhomogeneous Faddeev--type
equations. In \cite{ELO97,ELO97b,ELO97c} local NN (Malfliet--Tjon type) potentials
reproducing $s$--wave NN phase shifts up to the pion threshold were
used. Eq. (\ref{eq:equ}) was solved with the help of the Jastrow
correlated hyperspherical harmonic
basis (CHH).  The $\sigma_I$ values were taken from the interval 5
MeV$\leq\sigma_I\leq$ 20 MeV. The smaller $\sigma_I$ the more
accurate is the inversion but the larger is the number of CHH
 basis
functions required to achieve the same accuracy in the transform.
In fact for $\sigma_I$ tending to zero the scattering regime is 
recovered. In other
problems, such as the response of the $^{11}$Li nucleus in the
3--cluster model much lower $\sigma_I$ values can easily be used.
For an efficient inversion the transform should be obtained in the
$\sigma_R$ region covering the interval of $\epsilon$ values of
interest. Besides $\sigma_I=const$, the choice
$\sigma_I=\sigma_I(\sigma_R)$ is possible. In Refs. 
\cite{ELO97b,ELO97c} the $\sigma_I$ values
were taken to increase with $\sigma_R$. This improved the accuracy
in $R$ at low energy.

To judge the quality of the responses obtained the following criteria
were applied. First, the input transform should be obtained from the
dynamic equation (\ref{eq:equ}) with a sufficient accuracy. A typical
trend of convergence is demonstrated in Fig. 2. The quantity $K_{max}$ 
governs the number of the CHH basis functions retained
at solving Eq. (\ref{eq:equ}). The $K_{max}$ values in Fig. 2 are
sufficient to provide an accurate response and they
are substantially lower than those required in the bound--state
calculation with the same NN potential.
Second, as said above, the responses obtained should be stable with
respect to a change in $N$ in Eq. (\ref{eq:exp}) in some $\Delta N$
interval. This is the most important criterion. 
In the harder 4-nucleon calculations we had
$\Delta N\simeq 3-4$. In some easier cases the $\Delta N$ values 
were much larger. Third, various choices of $\sigma_I$ in Eq. 
(\ref{eq:ker}) and sets $\{\chi_n\}$ in Eq. (\ref{eq:exp}) should 
lead to approximately the same responses provided that the above 
mentioned criterion is fulfilled, and this also tested in our calculations as well. An additional check is provided
by sum rules. The quantities $\int R(\epsilon)d\epsilon$ and
$\int R(\epsilon)\epsilon\, d\epsilon$ were compared
with the sum rule expressions
evaluated independently. The differences were about 1\% or less for
the non--energy weighted and less than 2\% for the energy--weighted
sum rules.

To sum up, the method proved to be a practical tool for studying
few--nucleon responses. It would be of interest to try the approach 
also for exclusive and strong interaction induced reactions.

\begin{figure}
\caption{Longitudinal deutron form factor at $q^2=5 \;{\rm fm}^{-2}$. 
Conventional calculation: full curve; from inversion of the Lorentz transform: 
dashed curve.}   
\end{figure}

\begin{figure}
\caption{The Lorentz Transform with various $K_{max}$ values.}
\end{figure}

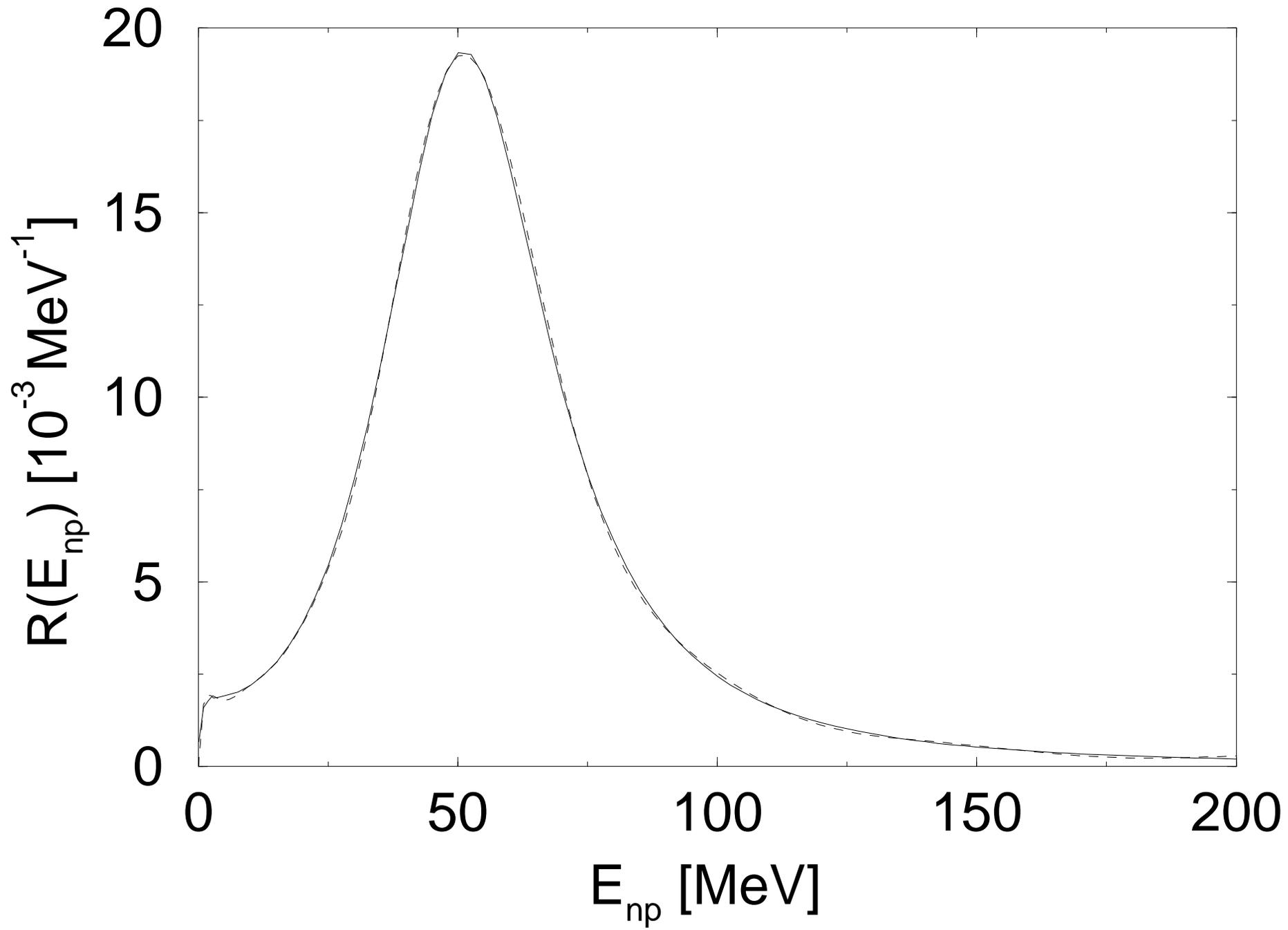
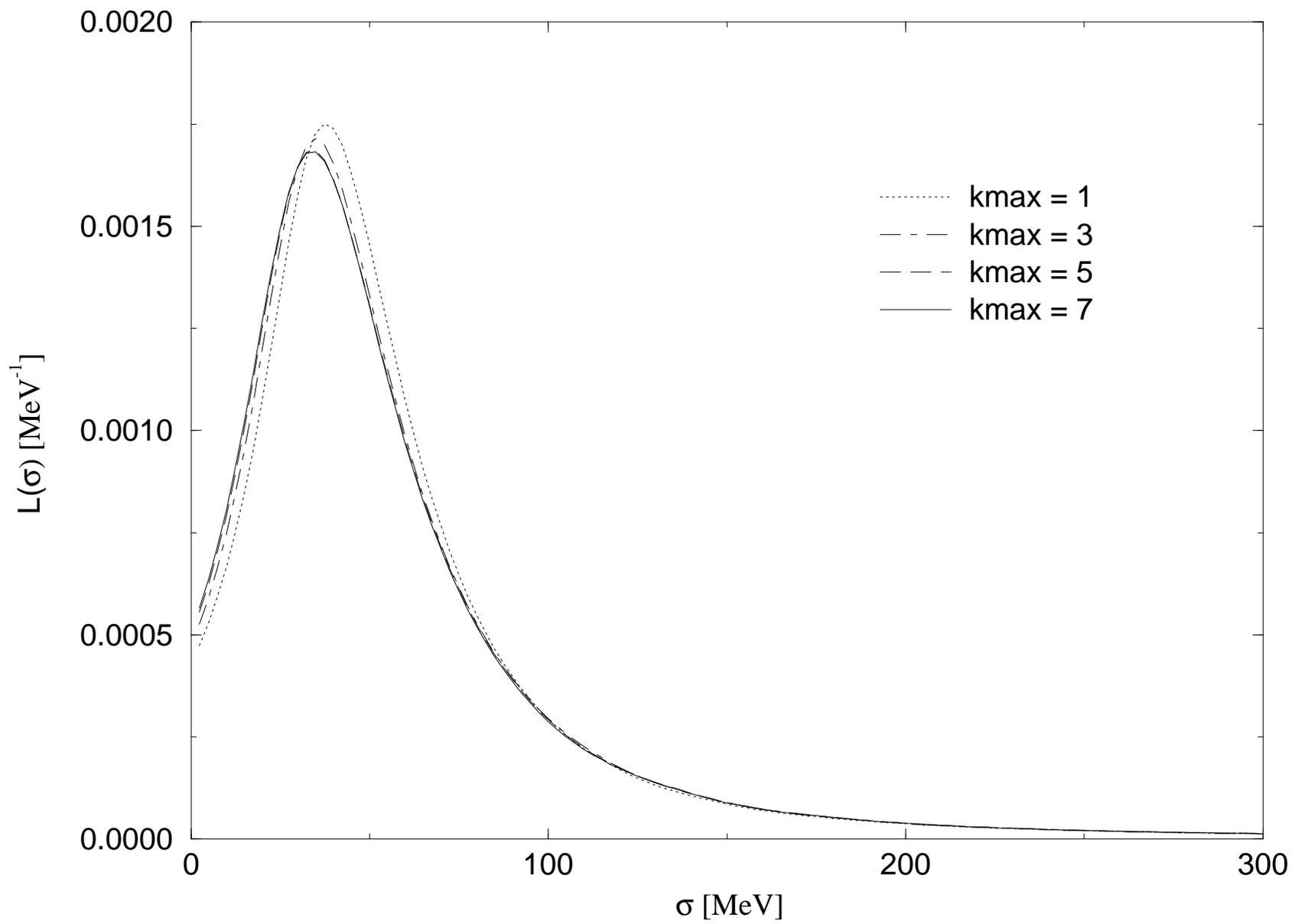
\end{document}